# Older Adults Imagining Future Technologies in Participatory Design Workshops: Supporting Continuity in the Pursuit of Meaningful Activities

Wei Zhao

The University of Melbourne, Australia, zhao.w3@unimelb.edu.au

Ryan M. Kelly

RMIT University, Australia, ryan.kelly@rmit.edu.au

Melissa J. Rogerson

The University of Melbourne, Australia, melissa.rogerson@unimelb.edu.au

Jenny Waycott

The University of Melbourne, Australia, jwaycott@unimelb.edu.au

Recent innovations in digital technology offer significant opportunities for older adults to engage in meaningful activities. To investigate older adults' perceptions of using existing and emerging technologies for meaningful activities, we conducted three participatory design workshops and follow-up interviews with adults aged over 65. The workshops encompassed discussions on existing technologies for meaningful activities, demonstrations of emerging technologies such as VR, AR, and AI, and design activities including prototyping and storyboarding. Our findings show that while participants had diverse interpretations of meaningful activities, they sought to use technologies to support continuity in the pursuit of these activities. Specifically, participants highlighted the importance of safe aging at home, which provides a pathway for meaningful activities in later life. We further discuss participants' discerning attitudes when assessing the use of different technologies for meaningful activities and several values and attributes they desire when envisioning future technologies, including simplicity, positivity, proactivity, and integration.

## 1 INTRODUCTION

In 1997, Rowe and Kahn defined *successful aging* as comprising three main components: low probability of disease-related disability, high cognitive and physical abilities, and meaningful engagement with life [79]. Over the past few decades, the advancement of digital technologies has revolutionised the lives of older adults, covering a spectrum from clinical innovation to shaping the everyday life experiences of individuals. Digital technologies are increasingly seen as essential for maintaining older adults' social connectedness [12, 88], social participation [7, 10], and engagement in meaningful activities [68, 94]. Older adults' spaces for meaningful engagement are not limited to the physical but are open to wider digital environments where they can learn, grow, socialise, and improve their quality of life [84, 93]. Through online platforms, older adults can take up new hobbies, learn new skills, connect with like-minded people, and participate in a range of local and far-flung activities [93].

Recent research in Human-Computer Interaction (HCI) has highlighted the value of engaging and empowering older adults in the design and use of creative technologies [58, 78]. Many studies have challenged stereotypes that position older adults as non-users of technologies or resistant to new technologies [13, 43]. Rather, engaging older adults in the design process can spark their creativity and wisdom and reflect the diverse range of skills they have gained through life experience, contributing to the invention of future technologies [71, 78]. Older adults have been invited to participate in, among other innovations, the design and use of virtual reality (VR) and 3D avatars [11], Internet of Things (IoT) systems [71], conversational agents such as Google Home and Alexa [95], healthcare systems [27, 39], and robots [3]. Yet, there is still limited investigation of how these emerging technologies can support or expand older adults' engagement in *meaningful activities* that bring value to people and help them achieve their goals or purposes in life, such as pursuing hobbies, volunteering, civic participation, and lifelong learning [37, 69, 93]. We see an opportunity to investigate older adults' interpretations of meaningful activities and perceptions of using different technologies to support these activities. Understanding these aspects can inform future technology design by provoking critical reflections on the benefits and limitations of digital technology for older people and identifying the values and attributes they desire when interacting with technology.

In this paper, we aim to examine how a group of independently living older adults: (1) perceive the use of *existing and emerging technologies* for supporting meaningful activities in later life, and (2) imagine *future technologies* that can support or expand their engagement in meaningful activities. To achieve this, we conducted participatory design workshops with 16 older adults aged 67-84 and living independently in their homes. The workshops involved focus group discussions, hands-on experience with technologies, brainstorming, prototyping, and storyboarding. We also conducted follow-up interviews with 10 workshop participants.

Our findings show that participants had various experiences of meaningful activities, including personal hobbies and interests, social interactions, physical activities, learning and enriching experiences, and entertainment and relaxation. In assessing how different technologies can be used to support these activities, participants expressed discerning attitudes towards when and how to use them and the potential harms of certain technologies. Participants' visions of future technologies revealed a shift in focus from meaningful activities to enabling safe aging at home, which was seen as a pathway for pursuing meaningful activities in later stages of life. Their design ideas revealed several values and attributes they desire in future technologies, including simplicity, positivity, proactivity, and integration. The findings were further interpreted through a lens of *Continuity Theory* [4], which suggests that people strive to maintain a consistent sense of self and a level of continuity in their lives, values, and activities as they adapt to the aging process.

Our study contributes to existing HCI research through a central argument that **older adults seek to retain continuity as they envision the future design of technology to support engagement in meaningful activities in later life**. This includes a desire to use technologies to maintain established patterns of meaningful activities, to expand and improve existing systems to create enhanced experiences, and to extend the continuity into the future when their capabilities are threatened by age-related factors. We argue that there is no one-size-fits-all design solution that can support older adults' engagement in meaningful activities. Based on these insights, we propose four lessons and opportunities. Future technology design and research for older adults could explore: 1) Designing for continuity in meaningful activities; 2) integrating social elements into IoT systems; 3) using AI as digital coaches for learning support; and 4) embracing the complexity of values in technology design.

## 2 BACKGROUND

### 2.1 Technology Design for Older Adults: Past and Present

There has been a long history in the HCI community of designing technology for older adults. Early work focused on age-related changes that have implications for the design of computer systems, such as reductions in vision [49], difficulties in using input devices due to arthritis or reduced spatial abilities [23], and declines in cognitive functions [24]. In recent decades, research on aging and technology has moved from designing for *deficiency* to a more *user-centred* approach, which means taking older adults' wants, needs, desires, and expectations into account as the underlying basis for design [60]. Researchers have used methods such as surveys, interviews, and focus groups to gather the needs and design requirements of older adults [21, 62] and evaluate prototypes in laboratory or home settings [29, 70].

There is a growing trend in HCI research towards a more proactive approach that focuses on *empowerment*, which often involves engaging older adults in the design and development process [9, 27, 33, 39, 71]. This is usually conducted through a co-design or participatory design (PD) approach, a cooperative method that treats those who will be affected by the new artefact as co-creators and co-designers [80]. For example, Rogers et al. [78] ran a series of workshops where groups of retired people used creative toolkits to invent the future and suggest ideas for new technologies. Their findings suggest that the value of this approach lies not only in working with participants to gather information but also in leveraging the creative abilities of older adults to inform future design opportunities. Similarly, Antony et al [3] conducted interviews and co-design workshops with older adults to empower older adults as drivers of the design of robots to engage them in physical activities. The authors identified several design guidelines that should be considered when designing robots for older adults, demonstrating the great value of co-designing with older adults to inform the development of new technologies. In our work, we actively involved older adult participants in PD workshops to aim for creativity and ensure that technologies are designed both *with* and *for* them.

### 2.2 Digital Technologies & Meaningful Engagement in Later Life

Rowe and Kahn [79] defined two main components of meaningful engagement for successful aging: *maintaining interpersonal relationships* and *productive activities*. Digital technologies have created significant opportunities for older people to stay connected to others over long distances and to participate in a variety of meaningful activities.

A great body of research in HCI related to older adults' interpersonal relationships focuses on maintaining feelings of connectedness at a distance [12, 18, 75, 88]. These usually involve promoting older adults' intergenerational communication with their grandchildren and adult children through technologies such as video calling, digital storytelling and shared tabletop systems [2, 73, 87, 92]. Some found that the intergenerational interaction is asymmetrical, with older people putting more effort into communication and compromising on their schedules more often than younger family members for fear of disrupting them [57, 67]. Another line of work focused on maintaining older adults' broader circles of contact and encouraging the development of new relationships through technologies [47, 54]. For example, Keyani et al. [47] built an augmented environment where older people could dance to clips from well-known films with others in groups, which helped them build new relationships with strangers. Digital technology can provide not just human-to-human relationships but a sense of companionship through social robots [e.g., 82]. However, it remains controversial whether these interactions contribute to meaningful engagement, as many people prefer to view technologies such as robots as assistants rather than companions [26].

Previous research has shown that some older adults seek communication with a level of dedication that cannot be supported through lightweight communication alone, which is often fleeting and dispersed over time [56]. They are more motivated to spend time on relationships that are emotionally rewarding and significant to them and prefer effortful communication that involves personalised and focused interaction with people [55, 56]. Building on this, Hope et al. [41] analysed the importance of paper-based and hand-written communications for older people. They argued that material social communication affords expression of thoughtfulness and concern, which suggests an opportunity to incorporate physical and material artefacts into online experiences. Other researchers have explored the role of digital technologies in supporting multimodal connectedness for older adults, navigating asynchronous and synchronous communication and using multiple communication channels for distanced interactions [35, 75]. These studies shed light on the meaningful aspects that some older people seek in their social interactions.

Another growing body of work in HCI focuses on sustaining productivity in later life through technology-mediated meaningful activities [7, 9, 74, 91, 93]. Some researchers focused on crafting and making as ways of creative production [42, 51, 74]. For example, Kalma et al. [42] explored the role of technology in a crafting group and identified opportunities to design for creativity, a sense of belonging, and quality relationships for older adult crafters. Richards et al. [74] explored some older adults' use of maker technology and uncovered opportunities for embedding small electronics in crafting. Other researchers have explored the use of video conferencing and social VR in older people's social participation practices [9, 93]. Zhao et al. [93] found that during the COVID-19 pandemic, some older adults used Zoom and other platforms to participate in various online social activities such as arts, sports, cultural, and civic events. They found that participating in these activities helped participants learn new skills, share care and support during difficult times, and gain entertainment and distraction. Baker et al. [9] collaborated with some older adults to explore the value of different social VR experiences. Their findings suggest that social VR can be used to provide shared experiences for older adults, help them reminisce about past experiences, and challenge societal stereotypes of aging.

Despite these varied explorations of how digital technologies can contribute to older adults' wellbeing, research efforts to date have not examined how older adults perceive the use of emerging forms of technologies such as artificial intelligence (AI) for pursuing meaningful activities. This paper seeks to address this gap by exploring a group of older adults' perceptions of various new and emerging technologies, particularly their potential uses for meaningful activities.

## 3 METHODS

Our research adopted an *exploratory*, *qualitative* approach to gain deep insights into older people's preferences for using technologies for meaningful activities. We organised a series of participatory design workshops to create a collaborative space for participants to contribute their ideas and engage in the design process. We conducted individual follow-up interviews with a subset of the participants to gather their additional comments on the topic. All procedures were approved by our university's Human Research Ethics Committee.

### 3.1 Participants

We recruited adults who were over 65 years old, living independently, had basic technology skills, and were able to travel to the study site (university campus) to participate in the workshops. We shared recruitment information through social media and posted recruitment flyers on information boards in local libraries. We also sent invitation emails to people who had participated in previous studies conducted by our research team. Our study focuses on the experiences of older adults living independently at home rather than those in aged care homes. The reason is that in Australia, where the study was conducted, living independently at home is the more common aging experience, with only a small percentage of people with high care needs going into aged care homes [6].

A total of 28 people expressed interest in participating, but due to travelling and time constraints, we finally had 16 participants (11 women and 5 men) agree to participate in the study. The participants were aged between 67 and 84 (M = 72.8, SD = 4.7 years). Seven were married and living with their spouse; nine were living alone, of which three were widowed, three were divorced, and three had never been married. When participants were asked about their ethnicity, most (13) said they were Australian, three said they were Asian, and one participant said they were born in Germany. We asked participants to describe their technological skills according to the Dreyfus five-stage model of adult skill acquisition [32]. Two participants identified their skills as novices; seven as advanced beginners; seven as competent; and none as proficient or expert. Appendix A.1 provides more details about the participants and their demographics.

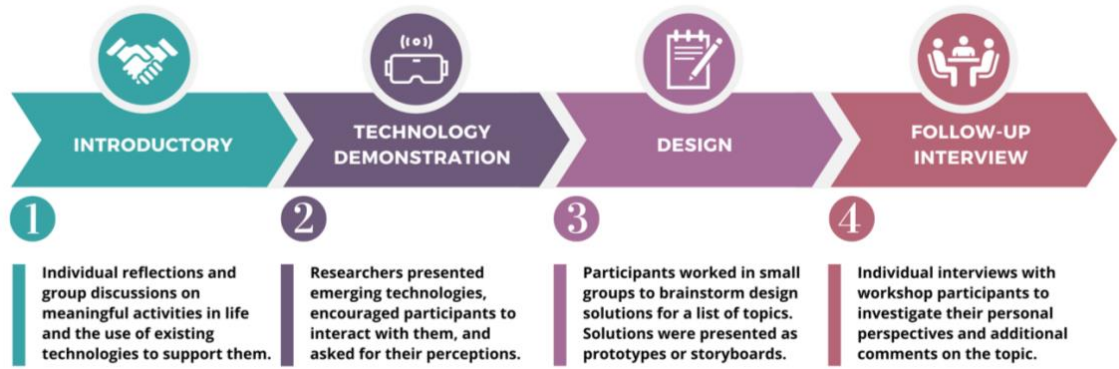

Figure 1: A summary of the four stages of the study.

### 3.2 Study Procedures

Figure 1 presents a summary of the four stages of the study, which started with a series of participatory design workshops. The workshops were organised by the researchers and held in conference rooms at the university campus. The first two workshops were held twice, with a subset of participants attending each one (Workshops 1a, 1b, 2a, 2b). We split the first two workshops to accommodate varying availability and ensure that everyone could participate meaningfully in the focus groups [65]. All participants were invited to attend the final design workshop held with the whole group (Workshop 3). Each workshop lasted between 90 and 120 minutes. Participants were provided with a $30 AUD gift card for each workshop they attended. We conducted follow-up interviews with participants two weeks after the workshops. The interviews were voluntary without compensation.

As the study spanned six weeks, participants' attendance during the workshops and interviews varied due to individual availability. Participant numbers for each phase of the study were: Workshop 1 - Introductory Session (n=15), Workshop 2 - Technology Demonstration Session (n=13), Workshop 3 - Design Session (n=12), and Follow-up Interviews (n=10). Three participants were unable to attend Workshop 2, and one participant was only able to attend Workshop 3 (see also Appendix A.1 for participants' individual attendance). This is common with multi-stage studies; for instance, in Baker et al.'s study [8], participants had different degrees of participation due to scheduling constraints and because some participants went on vacation during the study period. This has minimal effects on the results. In our study, the participant who only attended the last workshop may not have received as much information about the study as others, but he was able to follow the instructions and complete the workshop activities without difficulty.

**Workshop 1 – Introductory Session.** We first sought to have participants reflect on their perceptions of meaningful activities in their daily lives. After we introduced the objectives and plans of the study to the participants and all participants signed the consent form, we asked participants to write down activities they considered meaningful on individual post-it notes. At this stage, participants were told that they did not need to think about technology. This was intended to encourage participants to think more about themselves and less about designing for others, which was found to be common in other PD studies with older adults [71]. After that, we had a focus group discussion where participants took turns sharing the notes they had written, identified similarities and differences in their experiences of meaningful activities, and collectively discussed how existing technologies could be used to support these activities.

**Workshop 2 – Technology Demonstration Session.** We introduced participants to several emerging interactive technologies to gather their thoughts on whether these technologies can contribute to meaningful activities in later life. This workshop also aimed to provide inspiration for participants to prepare for the later workshop on envisioning future technologies. Participants took turns experiencing two head-mounted displays: an Oculus Quest 2 for virtual reality (VR) and a HoloLens 2 headset for augmented reality (AR) (see Figure 2). We prepared two types of VR experiences for the participants. One was a virtual tour of a city street of Barcelona using YouTube VR; the other was a social VR experience using VR Chat, where participants could walk along the beach as avatars and interact with others in the virtual space. For the AR experience, we used the Graffiti drawing application that allowed participants to draw lines in the air with their hands. Then, we introduced generative AI to participants and encouraged participants to pose questions to ChatGPT, a large language model developed by OpenAI. We introduced other technologies through videos, including the Amazon Echo Dot voice assistant [86], the Buddy companion robot [64], and the interactive table *Tovertafel* [38].

Additionally, we took participants on a guided tour of our *Interactive Technologies* laboratory. We described the functionalities of the equipment and devices in the lab, including motion capture tools, 360-degree projectors, 3D printers, a mini makerspace, and other ongoing research projects. We then facilitated a focus group discussion on how these emerging interactive technologies could be useful for supporting meaningful activities in later life.

**Workshop 3 – Design Session.** Before the session, we conducted an initial analysis of previous workshops to identify challenges and needs of participants in using technologies to support meaningful activities in later life. We identified a list of five design topics through affinity mapping. Participants formed four groups of three members, and each group selected a specific design topic from the given list. These groups were identified as Groups A, B, C, and D (see Appendix A.2 for the list of topics and group allocations).

Then, participants *brainstormed* designed solutions by writing down each new idea on an individual post-it note. They were encouraged to focus on quantity and come up with as many ideas as possible. We asked participants to vote for their favourite ideas and develop them further. After that, we guided the groups to create *prototypes* of their design ideas using the materials we provided, including large sheets of white paper, coloured paper, scissors, markers, coloured pencils, and oil pastels. Participants then used the *storyboard* templates we provided to illustrate how the design solution could be used in real-life scenarios. The workshop concluded with groups sharing their final prototypes and storyboards (see Figure 2 for relevant photos).

**Follow-up Interviews.** We invited all workshop participants to individual follow-up interviews, and 10 participants were eventually interviewed. The interviews focused on their reflections on the design workshops and their additional thoughts and comments on the research topic. We asked participants to share their experiences of using technology for meaningful activities in their daily lives. We also asked participants to describe their personal feelings about the emerging technologies we introduced to them during the workshop. The interviews lasted 20 to 30 minutes. All interviews were conducted remotely via Zoom and recorded with informed consent.

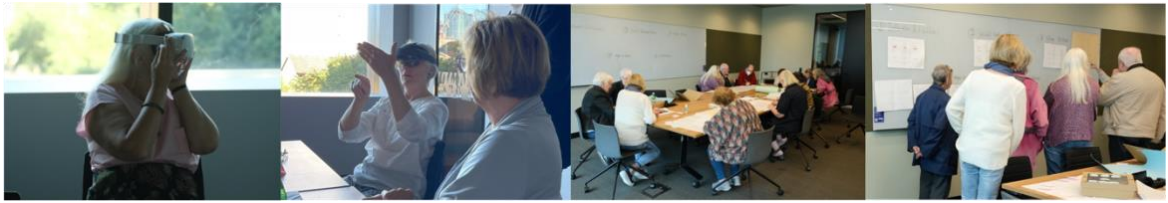

Figure 2: Photos of participants interacting with VR & AR headsets, engaging in group discussions, and presenting their design ideas.

### 3.3 Data Analysis

We collected a large amount of qualitative data in different forms from the workshops and interviews. Discussions during the workshops were audio-recorded and transcribed verbatim. We captured videos and photographs of key moments in the workshops, such as when participants were interacting with technologies and presenting their final design concepts. We also collected written and drawn materials from the workshops, including post-it notes participants had written about their personally meaningful activities in Workshop 1; the design ideas participants brainstormed in Workshop 3; and the final design prototypes and storyboards (Figures 3-7). The researchers wrote notes and memos during and after each workshop. Recordings of follow-up interviews were also transcribed.

The data analysis process was informed by the *reflexive thematic analysis* (TA) approach [16, 83]. The first author started the analysis by reviewing all the different types of data, focusing on translating them into written descriptions and documents. For example, the post-it notes and storyboards written by participants were collated into a written document for later analysis. Then, the first author conducted line-by-line coding of the data through an inductive approach. The codes were grouped into clusters, which were then organised into tentative themes. These tentative themes were written up as a summary document, which was then shared and discussed by the research team. Through iterative reviews and discussions, we dug deeper into the latent concepts underlying the data and refined the themes to represent the key insights more accurately. The themes were finalised and reviewed by the whole team. All authors were involved in the data analysis process and contributed to the analysis through their understanding of the topic.

## 4 FINDINGS

This section presents the main themes we identified from the study, including participants' diverse interpretations of meaningful activity, their discerning attitudes towards using technologies for meaningful activities, the shift of focus from meaningful activity to safe aging at home, and values and attributes participants desired in future technologies.

### 4.1 Diverse Interpretations of Meaningful Activity

We began the first workshop by asking participants to reflect on activities that they found meaningful in life. Participants had various interpretations of 'meaningful activity', and we categorised their explanations into five categories: 1) Personal hobbies and interests; 2) social interactions; 3) physical activities; 4) learning and enriching experiences; and 5) entertainment and relaxation. We further present how participants derived meanings from pursuing these activities.

**Personal Hobbies and Interests**. Participants described various hobbies and interests, including collecting stamps and coins; taking photos; painting; knitting and sewing; listening to podcasts, audiobooks, and music; dancing; watching

TV and films; reading; cooking; and gardening. Even for the same activity, participants derived different meanings from it. For example, P8 enjoyed gardening because she liked to watch things grow, while for P13, gardening was more of an activity that she "*had to do*" as a perceived obligation that was still meaningful to her. P15 saw gardening as "*a work of art*". P9 felt gardening was "*a meditative process*" for her, especially at the end of the day, but it could also be frustrating at times. These examples suggest that a meaningful activity may not always be enjoyable – the meaning an individual derives from an activity may come from the sense of achievement in completing it.

**Social Interactions**. A second category of meaningful activity centred on social interactions, such as "*visiting relatives*" (P3) and "*coffee time with friends*" (P8). Interacting with grandchildren was seen as particularly important and was described as "*inspiring*" (P12) and "*entertaining*" (P3). Participants also found attending social groups, clubs, and gatherings meaningful. For example, P11 described joining a walking group, which involved walking to different locations twice a week and having coffee with the group after each walk. The participant gained meaning from the activity through social support and a sense of community.

**Physical Activities**. Participants described different meaningful physical activities, including going to the gym, stretching, walking, and practising Pilates. These activities helped them maintain an active and healthy lifestyle, which was seen as the basis for enjoying other meaningful elements of life. For example, P2 said that his "*number one [meaningful activity] is interacting with my family and being active to enjoy it.*" Here, P2 considered it important to stay healthy, so he could be active enough to enjoy the social interactions that he cherished in life. Similarly, P4 noted that if a person wanted to visit a particular place, they would need transport to get there, which would require them to be physically fit to either catch public transport or to drive. This speaks to the importance of mobility for quality of life. Engaging in physical activities was also seen as vital for fostering social interactions. P6, for example, went to the gym every morning where she could talk to many young people. P9 found walking enjoyable for both her fitness and connections with others.

**Learning and Enriching Experiences**. Participants described various learning and enriching experiences they considered meaningful. These include attending workshops and discussion forums, participating in online courses, learning new languages, and exploring historical sites. Participants found meaning in these activities through skill development and personal growth. Some engaged in groups that offered learning opportunities; P13, for instance, enjoyed learning and acquiring new technology skills at the library, such as podcasting and downloading films to an iPad. P5 and P6 were members of the University of Third Age (U3A), which provides courses for retirees in different communities. Similarly, P2 participated in a group called *Sons of the West*, which offered lectures on topics such as health, nutrition, and wellbeing to older adults. These examples demonstrate that learning remains an important and meaningful activity that provides connections to knowledge about the world and local issues.

**Entertainment and Relaxation**. Participants described various playful and entertaining activities, including brain games such as puzzles, cryptic crosswords, Sudoku, and jigsaws; card games such as Bridge, Solitaire, Hearts, and Rummikub; and word games such as Wordle and Scrabble. Participants felt that these games provided pleasure and kept their minds active. For example, P9 played a brain-training game called Lumosity, which she felt was a good way for her to improve dexterity. Participants also expressed feelings of ambivalence towards games; for instance, P4 said about playing brain games on an iPad: "*They are wonderful. They are addictive and terribly time-consuming. I don't get up and exercise. I don't go to bed at night.*"

While most activities participants described were purposeful and intentional, there were other simple yet meaningful moments of relaxation in their daily lives. These included activities such as "*communing with nature*" (P4), "*sleeping in*" (P5), "*looking at the ocean from my window*" (P5), "*watching how dogs behave*" (P5), and "*eating anything*" (P6). These moments provided a sense of tranquillity and contentment amid everyday life.

From the above, meaningful activity is a very *individualised* concept. A meaningful activity can be individual or group-based, productive or unproductive, enjoyable or occasionally burdensome. These varied perspectives on what constitutes a meaningful activity set the stage for discussions in subsequent workshops.

### 4.2 Discerning Attitudes Towards Using Technologies for Meaningful Activities

In Workshops 1 and 2, we discussed with participants how existing and emerging technologies can be used to support the various meaningful activities mentioned above. We also encouraged participants to share their perceptions of the technologies during the follow-up interviews. Participants showed discerning attitudes towards using technologies for meaningful activities – seeing potential in some but remaining sceptical about others. Rather than simply accepting things at face value, they took time to analyse and consider the impact of the technologies on older people.

When reflecting on the possible roles of technology in supporting meaningful activity, participants initially discussed their experiences of the COVID-19 pandemic and how the technologies we presented in the workshop would have been helpful during the lockdowns. The city in which this study was conducted experienced one of the longest lockdowns in the world, totalling over 260 days as part of the COVID-19 pandemic control strategy [20]. A strict stay-at-home order was imposed, so people had only limited reasons to go outside. By the time we conducted the research, the strict restrictions had been eased, and people were gradually returning to their regular lives.

In adapting to the changes in different stages of the pandemic, participants in our study saw the advantages of using online tools for meaningful activities and expressed a desire to **have both online and face-to-face options**. Some participants described participating in various online social group activities during the restrictions, such as attending "*Zumba classes via Zoom*" (P6). As the restrictions were gradually eased, some of these activities were continued online. P14 said: "*We used to go for monthly lectures, and they never went back to face-to-face. It's been online ever since.*" Many participants enjoyed the convenience provided by these online opportunities, especially under special circumstances. As P4 stated, "*There is no way I'm going to do yoga in the park when it's raining or 10 degrees*". P5 said, "*It's helpful for people who might have difficulty getting to the physical location*".

However, other activities went back to face-to-face and ceased to be offered online after the lockdown. Many participants expressed disappointment with reduced availability of online options. For instance, P15 enjoyed taking Bollywood and Samba dance classes online, but she had to stop and look for alternatives when the classes moved back to face-to-face. She said: "*There was a lot of stuff online, and now that's stopped. You can't pick it up from the US because their timeframe is too difficult*". Some participants looked for activities on other online platforms such as YouTube, but they expressed concerns about the credentials of the YouTubers compared to scheduled classes on Zoom, especially regarding safety when doing exercises. Participants expressed a strong desire to have "*a mix of both*" (P6, P8), indicating their preference for having both online and face-to-face options available.

When reflecting on the use of current technologies for meaningful activities, participants had **varied thoughts on the value and impact of certain technologies**. Social media, in particular, received polarised comments from participants. For example, P2 was opposed to TikTok as he felt that TikTok users were "*getting out of touch with the reality of what's happening in the world*". He also felt that TikTok was being misused to portray fake content and convince gullible users that they were real. Contrarily, P4 enjoyed watching the videos appearing on TikTok's home page as this helped her find connections with the younger generation. This mixed feeling was mirrored on Facebook. Some participants enjoyed using it to stay in touch with their adult children and grandchildren, while others felt annoyed or even "*driven insane*" (P3) by the trivial or promotional content shared on it. Some also expressed concerns about privacy and security issues when using those social media platforms. These examples demonstrate participants' critical thoughts on the strengths and potential issues when making decisions about whether to use certain technologies.

Participants were also discerning when **assessing the usefulness and potential harms of the emerging technologies** we introduced to them. They actively imagined integrating these technologies into their daily lives or the lives of other older people, but at the same time questioned whether they were truly meaningful and beneficial. All participants experienced VR for the first time in our workshop. Most of them were amazed by the immersive features of VR; for instance, after the virtual tour, P8 exclaimed, "*It's like dropping into a new world!*". P8 explained that despite her interest in history and travel, she could not travel as much as she would like. She felt excited that technology like VR could enable her to continue to pursue her passion for travel. Beyond this application, other participants mentioned the possibilities of using VR to calm people down in stressful situations, bring back someone's memories of past lives, and facilitate engagement in artistic events.

During the discussion, participants expressed conflicted opinions on the value of social VR for addressing loneliness experienced by older people. Some believed that social VR could create a sense of community for lonely people, keeping them occupied and engaged. P13, for instance, felt that social VR could be useful for those not confident in face-to-face conversations, while P6 saw an opportunity for some people to use social VR for making new contacts and engage in online dating. However, other participants did not see it as meaningful social interaction and expressed concerns that it might exacerbate feelings of isolation:

> P7: "It would only be interaction if it was continuing, if you interacted with some people who you subsequently interacted with again and again."
> P11: "I think if I use this all the time, I will feel isolated. I need to talk to people face to face, and I can see the emotions when I talk [face to face]. And this [VR] is just a voice for me."
> P9: "There is tactile. There's something special about human connection."

When we asked participants to interact with ChatGPT, they were surprised at how quickly the responses were generated by AI and the amount of information it could provide. One participant asked ChatGPT to write a story about 'murder in a village'. In less than a minute, a story was born about the investigation process of the murder of a local baker, Mr Smith, and how the truth gradually came to light. Participants found the story very interesting, with one saying: "*I'd like to read that book!*". Through discussion, participants ideated different ways AI might be useful in their lives. They particularly focused on using AI as a personal assistant for solving technical issues, such as using it to convert PDF files to Word documents (P3), download photos from a mobile phone to a computer (P2), set up a new laptop (P13), simplify expert-written content into layman's terms (P13), and translate content into different languages (P11). However, some participants expressed criticism that such generative AI tools might make people more mentally lazy. P11 said: *"One day maybe we don't need to think anymore; somebody else is thinking for me."* Some also raised concerns about the credibility of the results, as P14 stated: "*AI is obviously programmed according to people's prejudices*" (P14).

These examples show that our participants were actively engaged in discussions about technology. Participants were carefully **making informed choices about what technologies mean to older people, the situations of using them, and the potential harm the technologies could do to certain populations**. We believe that these discerning attitudes they expressed towards technology affected their design decisions in the subsequent design workshop, particularly in terms of considering how and why to use a certain technology and its potential impact on users.

### 4.3 Preparing for Older Age: From Meaningful Activity to Safe Aging at Home

As the discussion about meaningful activity and technologies converged, we identified a shift in focus towards using technologies for enabling safe aging at home, which then facilitates meaningful activities in later life. This was particularly evident in Workshop 3 where participants started to envisage future technologies.

Some design ideas were consistent with our initial expectation – designing systems to support current meaningful activities or improving existing systems to enhance their experiences. For example, Group A designed a smart gardening app called **See How Your Garden Grows** (see Figure 3) to support their interest in gardening. They envisaged users simulating the layout of a garden and then using the app to add or remove plants, adjust sizes, and experiment with different arrangements. Users can see how specific plants will grow over seasons and years and explore the garden from different angles. These features show how participants imagined future technologies to support their *existing interests and hobbies*.

A design idea from Group C focused on *improving existing systems to enhance experiences* when engaging in online activities from home. They designed an advanced video conferencing system called **Active Together Through Zoom** (see Figure 4). The system offers improved audio and visual performance compared to Zoom. It was envisaged that users can participate in group activities such as choir singing and educational seminars via video calls or VR headsets. Exercise and activity sessions are managed by human hosts with support from AI as co-hosts. First-time users receive an initial remote health assessment by an experienced physiotherapist; they will then receive personalised exercise plans, which may include yoga, cardio, and weight exercises. The system is integrated with a smart yoga mat with built-in sensors and a thermal camera. This allows users to get feedback from the system about their workout or exercise performance.

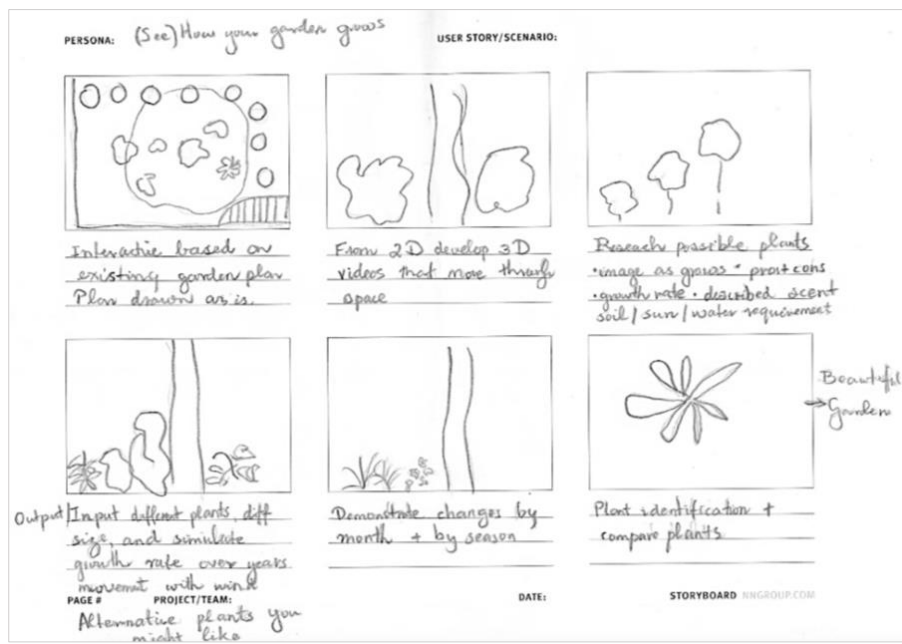

Figure 3: Participants' design of a smart gardening app *See How Your Garden Grows*. Users can use it to plan and customise their gardens and compare the pros and cons of different plants. It also visualises how a plant grows over seasons.

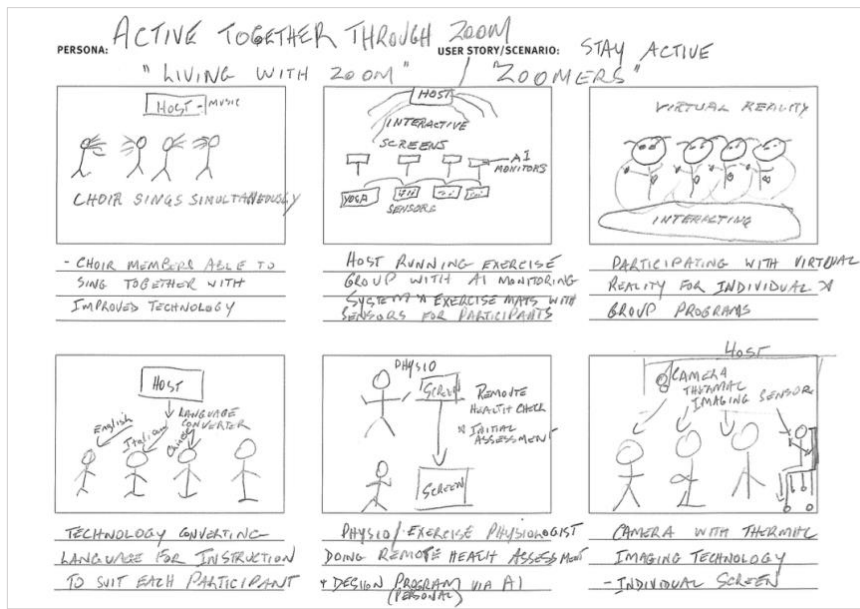

Figure 4: Participants' design of a video conferencing system *Active Together Through Zoom*. Users can participate in social activities through the system. It integrates AI as co-hosts of meetings to develop personalised exercise programs for users.

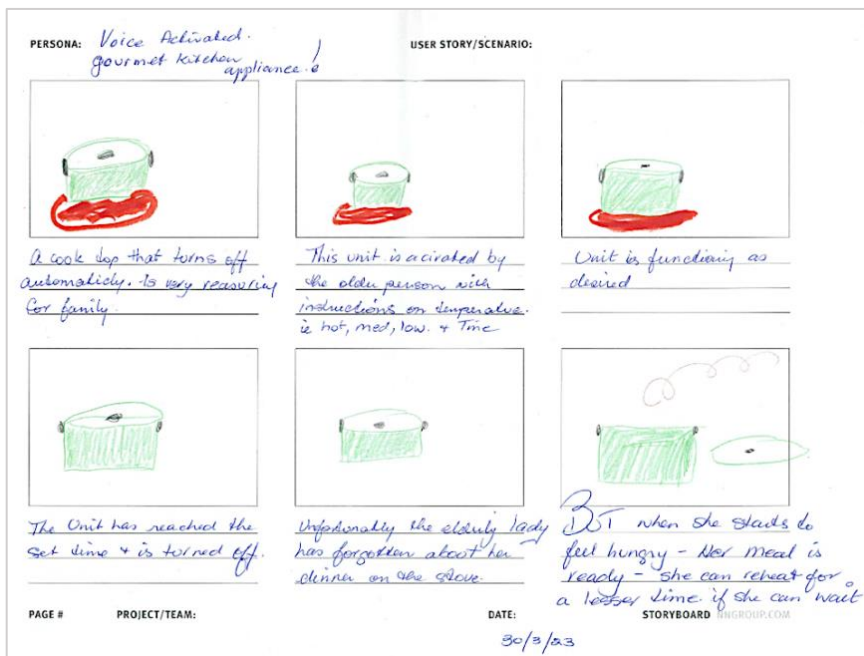

Figure 5: Participants' design of a smart cooktop system *Voice Activated Gourmet Kitchen Appliance*. Users can activate the system through voice commands such as cooking temperature and time. The system is envisaged to turn off automatically after a set time. Users could also reheat the food in a shorter time.

The intended purposes of other designs from participants, however, were different from what we initially envisaged. In these designs, the focus shifted from supporting meaningful activities to *enabling safe aging at home*. Through an analysis of those designs, we saw safe aging as a necessary precursor for our participants to continue their engagement in meaningful activities when they are threatened by age-related declines. For example, Group A designed a smart cooktop appliance called **Voice Activated Gourmet Kitchen Appliance**, which provides special features such as ingredient customisation, nutrition information display, reminders, and time-saving options (see Figure 5). Users can set the temperature and cooking time, after which the system will turn off automatically. The system is controlled by voice commands. P6 described how an older person living with dementia might benefit from the system:

> P6 (age 74): "My old lady (a fictional persona) who's got a bit of dementia put her dinner on, and she's got this fancy stove that looks fabulous, and she said to it: 'Please cook this casserole for 40 minutes on a medium heat.' And then she forgot all about it, and sometime much later, when she got hungry, she thought, 'Oh!' [She] went back into the kitchen, and there was her dinner cooked. There you go."

This scenario shows that participants envisaged the smart appliance would allow older people experiencing memory or concentration issues to use it without worrying about potential safety hazards. It also provides opportunities for those who enjoy cooking to pursue the hobby safely through periods of cognitive decline.

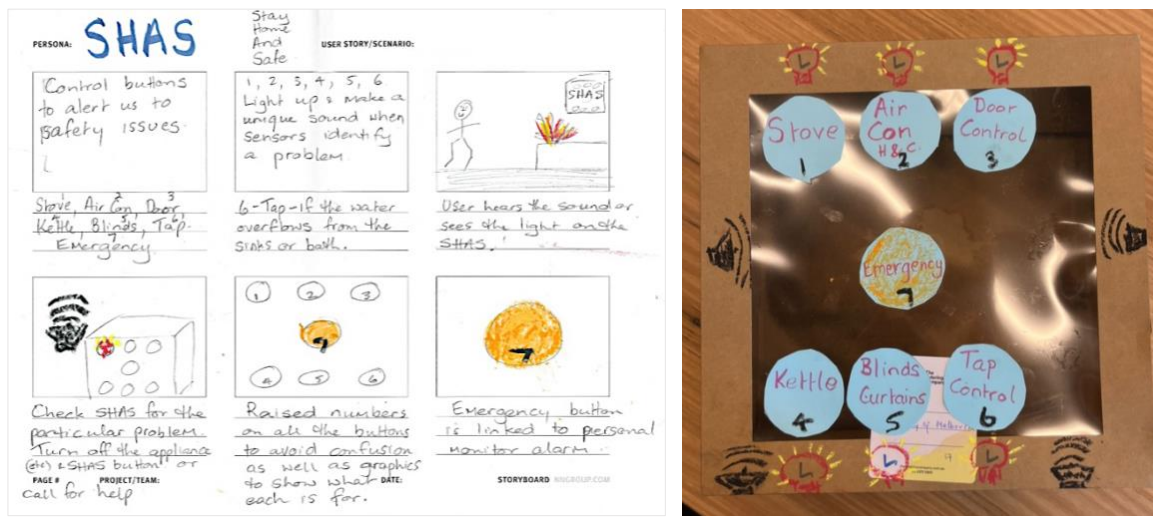

Figure 6: Participants' design of a home control system called *SHAS*. Users could control home appliances through a box-shaped panel, receive alerts when potential safety concerns are detected, and use the system to call emergency services.

Similarly, Group D designed a home control system called **Stay Home and Safe (SHAS)**. The system features a box-shaped control panel envisaged to be placed in the centre of an older person's home. The panel includes seven buttons used to control different components: Stove, Air Con, Door Control, Kettle, Blinds/Curtains, Tap Control, and Emergency.

> P9: "SHAS has a control centre which alerts our elderly person who's become a little bit vague and forgetful whenever they left the stove on, or the air conditioning needs attention. If you've left a tap on and then the water starts to overflow, you need an alarm to let us know, an emergency when things go completely pear-shaped. When there's a fire or that sort of thing, it goes to the local emergency services."

When we asked about the design intentions, participants mentioned that the relief from potential safety hazards at home would help older people to remain independent in their homes, so they could stay longer in their homes without having to move into a residential aged care facility.

In addition to safety alerts, participants designed different ways that older people could *control the system*. For instance, the SHAS system was designed with raised buttons in different shapes and colours for easier control, especially for those with speech impediments. As one participant explained, "*Sometimes they can't understand; they can touch rather than speak different commands.*" The system also provides strong visual feedback through blinking; one participant had hearing problems, so this feature was designed to cater to older people with similar needs.

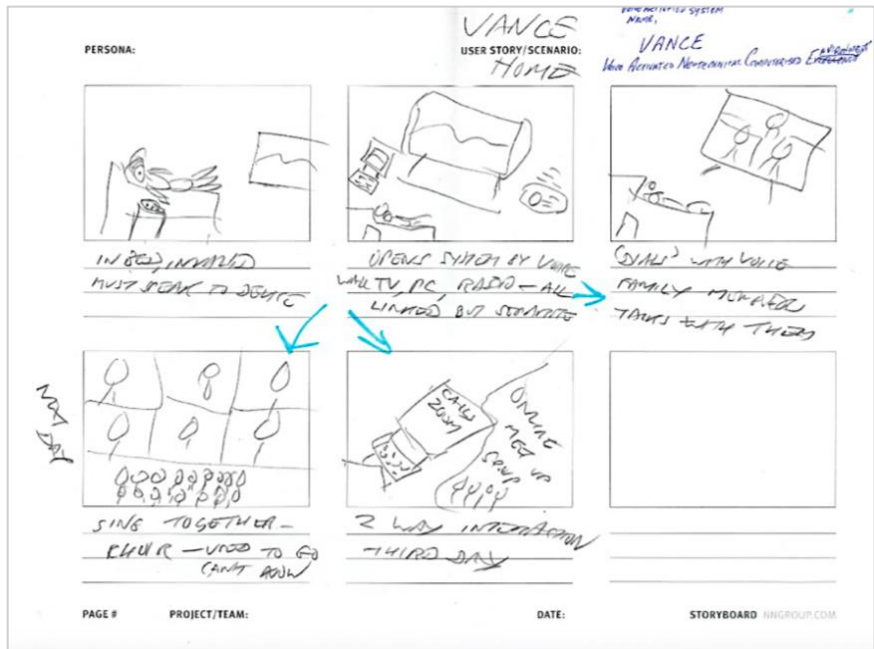

Figure 7: Participants' design of a voice-based home control system *VANCE*. Users could use it to control utilities such as televisions, radios, and personal computers. Users could also use it to call their family members and participate in online social activities.

Interestingly, while there was little discussion about voice interaction in the first two workshops, *voice control* became a major feature of most design concepts in the design workshop. For example, Group B designed a centralised home control system called **Voice Activated Neotechnical Computerised Environment (VANCE),** which is activated and operated through voice. Users can use it to control home utilities such as televisions, radios, and even personal computers. Users can also make calls and participate in group activities and meet-ups remotely. P16 believed that "*with less dexterity coming probably, maybe the voice is the way to go*", which speaks to the design goal of preserving the ability to control.

P16 (age 71): "I think probably a lot of it was thinking about where I might be in if I'm still around in 20 years' time. At the moment, I'm still very physically active. But there are things that are starting to impact on that.

What happens when my arthritis has progressed to the point where I can't use a keyboard or a phone? How do I continue to make sure that I'm able to maintain what I'm trying to do? I guess because so much of what I now do has to do with voice, it was voice that was the obvious solution. That's not going to be the solution for some people. And some people aren't going to be comfortable with that either."

P16 thought voice was the *"obvious solution"* because of his work in a radio station as a broadcaster. As he and other participants noted, while voice-activated systems are helpful for people with visual impairments and reduced dexterity, they may also pose challenges for those who have issues with speech or hearing. We believe that participants' interest in adding voice control to future technologies indicates an intention to prepare for older age and to continue engaging in meaningful activities by maintaining control and enabling safe aging.

### 4.4 Envisioning the Future: Desired Values and Attributes in Technology

Based on the above design concepts and discussions with participants in the follow-up interviews, we identified four nuanced values and attributes participants desired in future technologies: simplicity, positivity, proactivity, and integration. These values and attributes point towards how future technology could be better designed to support older adults' needs for pursuing meaningful activities through technology.

**Simplicity.** Having a simple, easy-to-use interface is participants' most desired attribute in future technology. Participants complained that many current digital systems require complex navigation, and they have to "*go through too many screens to get to something*" (P5). P6 described her feeling of frustration when technology was over-complicated:

P6 (age 74): "Don't you think that some developers have got a bit of ego involved and made things a little bit trickier so that they stand out? I sometimes try to do something complicated, and I'll have to get one of my boys to help me because it's just too hard. I'm not silly, but sometimes I just can't be bothered with the hoops they'll get you to jump through… Things have to be really mother-friendly, and really easy so that your five-year-old grandson doesn't have to come visit and help nana sort this out. That's what we need."

When we asked participants how technology could be designed to be simpler for older adults, some said that they expected future technologies to be more "*streamlined to make it clear*" (P14). This means that interfaces should be designed with minimal distractions and avoid intricate back-and-forth commands. Participants also desired fewer changes, and they wanted the "*old things to continue to work*"; they did not like having to adapt to a new interface every time a system was updated. For instance, P1 felt frustrated about the upgrades in the Windows system: "*One more ridiculous thing is, every month the computer is upgrading automatically, and then something is missing.*" Participants also expressed dissatisfaction about having to update certain mobile applications before they could continue to use them.

Additionally, participants expressed a need to **improve the instructions and manuals for technology**. Some complained that their limited experience with technology made it difficult for them to understand the technical terms used in the instructions. P4 said: "*Most of the instructions are foreign languages to me. I can't understand what they are talking about. I'm not that dumb, just inexperienced.*" P8 found it hard to find useful information on the help pages provided on websites. P7 struggled with understanding graphic and visual instructions and preferred to have step-by-step textual ones. These examples suggest participants' need for *scaffolded support* when using technologies.

Based on this, participants brainstormed ways for future technologies to provide better instructions for older adult users. P3 suggested attaching a video tutorial to the system to explain how to use the system. P4 recommended that future technologies should "*periodically have some sort of revisions on how to use it*". Some participants further envisaged the integration of AI assistants in digital systems to introduce users to the main features and help with troubleshooting. These suggestions present opportunities for designing simplified digital systems and providing scaffolded support.

**Positivity.** Participants highlighted the importance of having positive feedback from technologies for older users. In fact, many participants were positive about using new technologies for meaningful activities. They were willing to explore new technologies as long as they could provide positive feedback and were not getting "*too sophisticated*".

> P8 (age 67): "I guess I'm a great believer in that anything that makes life easier and more adaptable for me has got to be a good thing. I don't want to end up in that sort of fearful place of being frightened to use technology. I think that was one of the thoughts that came through from a number of people in the group I was in."

The positivity of digital technology is closely related to simplicity. Users often receive positive feedback when they complete their tasks and goals with minimum difficulty. Conversely, participants felt that the error messages displayed on some systems were too "*overwhelming and intimidating*". Participants expected that future technology would avoid these negative messages so that they do not have to worry about "*pushing the wrong button.*" Delivering positive messages in digital systems could support older adults' confidence in using and adapting to new technologies.

**Proactivity.** Participants envisioned future technologies to be more proactive rather than reactive. This includes proactively detecting safety risks and sending alerts to users. For example, the SHAS system was envisaged to send alerts to users when it detects anything unusual in the home. These reminders and alerts are particularly helpful for those who have difficulty remembering things, providing a sense of reassurance. The proactivity also lies in the delivery of information; for instance, the smart gardening app allows users to compare different plants by listing the pros and cons in real-time. Similarly, the advanced video conferencing system uses thermal cameras and sensors to actively monitor users' movements and provide real-time feedback to users. These examples indicate an **active feeding of information rather than waiting for user commands.**

Furthermore, while most groups in Workshop 3 added a voice activation function to their design concepts, some participants were not satisfied with existing voice assistants. While P16 advocated for voice interfaces in their group's design, he felt that existing voice assistants were very "*disembodied and mechanical*". They could only understand a limited number of commands. For most user input, they just come back and say, "*I don't understand*". Participants envisioned future voice assistants to be more *advanced*. For the VANCE system, participants imagined it to be smarter than existing voice assistants in that it would understand users' intentions and provide responses that are highly aligned with those intentions. As these examples show, participants expect future technologies to **proactively communicate with users, ask for their needs, adapt to different speech patterns and provide more accurate responses**.

**Integration.** Participants expected future technologies to integrate different technologies or functions into one system, especially technologies to be used in the home. For example, the advanced video conferencing system integrates technologies such as video conferencing, virtual reality, artificial intelligence, sensors, and thermal imaging. The SHAS system uses a box to represent a centralised panel that controls different smart components in the home, such as stoves, taps, blinds, and doors. P9 explained the inspiration behind this design:

> P9 (age 73): "The concept comes from very old English homes that used to have the maids, the providers of service, and the kitchen staff and the floors. And they have a system on the wall, which is a bell that's linked to different rooms, and it rings. It's derived from that, and we've got a facility that could happen in different rooms or for different pieces of equipment."

While we had limited information about participants' familiarity with the concept of IoT, this integrated design is consistent with the main characteristics of an IoT system. This consistency can also be found in the VANCE system. It emphasises the integration of multiple functions: "*Different things can talk to each other, and it's all in one system*" (P16). Participants valued and prioritised **the power of control, ease of use, convenience, and seamless interaction** in these integrated systems, which are important in their use of technologies for meaningful activities.

## 5 DISCUSSION

This research examines how a group of independently living older adults perceive the use of existing and emerging technologies and imagine future technologies for meaningful activities in later life. Our study extends existing research on designing technologies to support the social connection and the psychosocial wellbeing of older adults [11, 78].

Our findings shed light on diverse interpretations of the notion of 'meaningful activity'. People are interested in different things and have diverse social lives, so they have a variety of activities and people who give their life meaning. This diversity poses a challenge for design work – **there is no one-size-fits-all design solution that can support engagement in meaningful activities for every older adult**. Recent design work for older people has increasingly shifted towards a person-centred approach, with a focus on designing for situated communities [76]. In our initial analysis, while we saw participants' perceptions of meaningful activities as personal and individual, we found commonalities in their motivations and intentions for engaging in these activities. Notably, we identified a shared desire to stay active, foster meaningful social interactions, and promote personal growth. Participants also preferred activities that were creative, playful, educational, and relaxing. Instead of designing for individual activities, we argue that designing technologies for meaningful activities is about recognising the meanings and values that older people seek from the activities and addressing the challenges they encounter when pursuing the activities.

We also presented participants' discerning attitudes when assessing the use of existing and emerging technologies for meaningful activities. They embraced the use of new technologies in their lives but reflected thoughtfully on the potential harms and issues new technologies might cause. These attitudes affected their design decisions in the workshop where they envisioned future technologies. They made informed choices about how and why to incorporate certain technologies and what impact the design might have on older people's lives.

Participants' design ideas emphasise an underlying need to enable safety, autonomy, and independence in the home while aging. These needs align with previous research on aging in place [e.g., 45, 50]. In our study, supporting these needs through technology was seen as a necessary pathway for continuing meaningful activities as people began to face age-related decline. This offers a new perspective for understanding older adults' preferences for aging in place.

In the following section, we present our interpretations of the findings through the lens of *Continuity Theory*. Drawing on theory can establish common terminology, explain and contextualise research, inform design practices, and generate questions for further study [22, 77]. Our analysis suggests that **older adults seek to retain continuity as they envision the future design of technology to support engagement in meaningful activities in later life**. We then discuss the lessons we learned from the study and the implications of this research for future design and research practices.

### 5.1 Continuity, Adaptation, and Later Life

The Continuity Theory of Normal Aging [4] proposes that as people age, they tend to preserve and maintain existing psychological and social patterns by applying familiar knowledge, skills and strategies. In the context of activities, continuity theory suggests that adults develop stable *patterns of activity* over time and take measures to maintain these patterns in adapting to aging [5]. The motivation for retaining continuity in activities is heightened when people begin to stake their identities on an activity; for example, in the case of a seventy-year-old with a forty-year history of interest in cars or music, the interest may represent a central theme in life and be used to define the person [34]. Continuity theory also posits that individuals make choices not only to achieve their goals but also to adapt to changing environments [46]. Thus, when faced with adaptive challenges, people tend to rely on what they perceive as their *established adaptive strengths* [46]. Applying this theory to our findings reveals three areas of continuity and adaptation that older adults seek to maintain in their use of technologies for meaningful activities: continuity in activity engagement, continuity in technology competence, and adaptation to changing contexts.

**Continuity in Activity Engagement.** We started this study by exploring how technologies could be designed to support and expand older adults' engagement in meaningful activities. Some designs were consistent with our initial expectations. For example, participants designed the smart gardening app (Figure 3) to support their existing interest in gardening, and the advanced video conferencing system (Figure 4) was envisaged to improve participants' current experiences of using Zoom. These examples suggest a need for technologies to support **continuity in pursuing existing patterns of meaningful activities** and enhance the experience.

Other designs, however, focused more on using technologies to enable safe aging at home. Participants envisioned technologies they might need in later stages when confronted with age-related threats and challenges. Through these designs, participants expressed a desire for continuity in their capacity to engage in meaningful activities and extend that continuity to the future. For example, the smart cooktop system (Figure 5) and the use of voice interaction in most designs show that participants wanted to **maintain the capability to control things** when they might experience cognitive decline or limited manual dexterity.

In addition, participants desired technology that would help them stay in their homes longer and reduce the need for care so that they would not have to move into a residential aged care home. Residents of aged care homes are usually expected to spend their time on a predetermined schedule with limited autonomy [85]. Transitioning to care homes presents challenges such as loss of contact with local communities and reduced flexibility in time management. The desire to remain at home demonstrates a need for **continuity in the opportunity and autonomy** to engage in meaningful activities as people age.

**Continuity in Technology Competence.** Consistent with prior research [94], participants in our study complained about the frequent updates of systems and applications and the inconsistency between versions. They wanted "*old things to continue to work*" so they could maintain the knowledge and technical skills they had mastered. This is also related to older adults' confidence in learning to use technologies and their need for competence through achievement [94]. For those who are unfamiliar with digital interfaces and require more time to learn how to navigate a system, each significant interface update necessitates a complete relearning and adaptation process for the user.

However, it is important to note that **the desire for continuity does not mean resistance to change**. In our analysis, participants were actively involved in the discussions about emerging technologies and brainstorming innovative ways to use technologies for meaningful activities. Atchley [5] highlighted the paradox of continuity in adult development, where there is both consistency over time and obvious changes. Individuals can exhibit similarities when compared to their past selves while undergoing significant changes in the process. Our analysis reveals that older adults seek to retain continuity in the established skills and knowledge they have mastered through learning, while at the same time embracing new technologies that could be beneficial to their lives.

**Adaptation to Changing Contexts**. In later adulthood, individuals may experience many transitions or life events that can change the context in which they engage in meaningful activities, such as retirement, widowhood, and empty nests [25]. The recent experience of the COVID-19 pandemic and the resulting restrictions created a stage of transition for many people, including older adults. In our study, participants described using Zoom and other online tools to participate in various social activities. This is consistent with previous studies that explored older adults' use of technologies for social connections during the pandemic [81, 93]. Evidence from prior studies suggests that many older adults used technology to find alternative options that helped them maintain the continuity of participation in meaningful activities in the changing context. Adding to their findings, our study discussed some older adults' experiences after the transition back to in-person activities. Through the online opportunities that emerged from the pandemic, our participants developed adaptive strengths and skills in remote communication and participation. They did not want to lose the convenience and diversity of options that were available online. In this case, the use of online tools for participating in social activities has become a new activity pattern that many older adults want to maintain, indicating a need for continuity in not only established skills and practices but also more recent behaviours.

**5.2 Future Technology Design Opportunities for Older Adults**

In this research, we adopted a participatory design approach where we explored the needs of target users not only through verbal descriptions but also through tangible design artefacts such as prototypes and storyboards. These artefacts enabled us to uncover practical insights about how older adults imagine future technologies to be used in real-life scenarios. Drawing on these insights, we present the following lessons and opportunities for future technology design to support older adults' pursuit of meaningful activities (Figure 8).

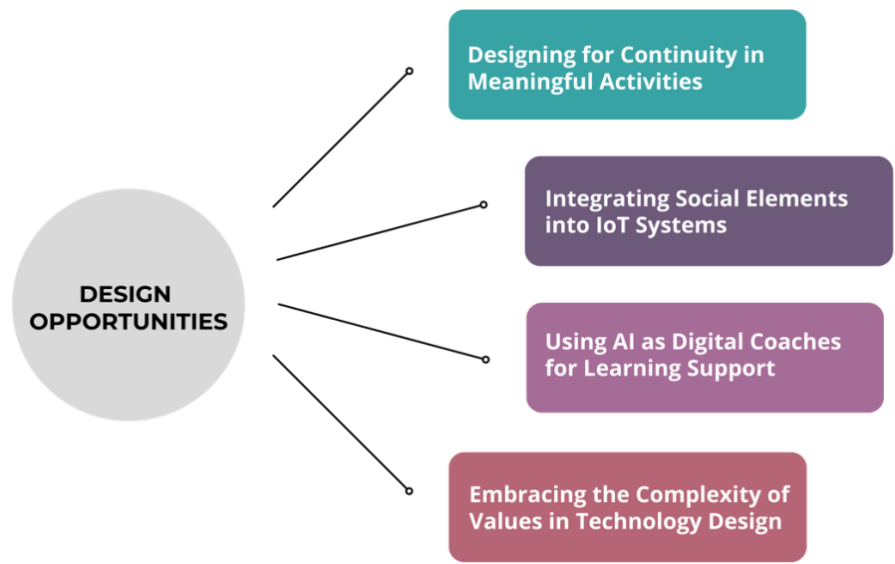

Figure 8: Four design opportunities and lessons for older adults' pursuit of meaningful activities

**Designing for Continuity in Meaningful Activities.** Our analysis shows that older adults seek to use technologies to retain continuity in the patterns of activities they engage in, the capacity to engage in the activities, and the opportunity and autonomy to engage in the activities. They also need to maintain continuity in their competence with using technologies and their adaptive skills. Drawing on these findings, we identify the following lessons on how future technologies can be designed to support older adults' continuity in meaningful activities:

- Designing technologies that provide *diverse activity options* for older adults. As meaningful activities are highly individualised, having access to technologies that can provide a wide spectrum of activity options is important. Video platforms such as YouTube and TikTok have shown their strengths, but how to meaningfully leverage other technologies to support a variety of activities remains to be explored.
- Using technologies to *replicate meaningful experiences* to counter age-related decline. Emerging technologies can replicate experiences that are meaningful but challenging for people with declining abilities. For example, VR could allow someone to relive a past interest in Formula 1 racing [90]. Future work should investigate how these experiences could be enhanced to support a consistent sense of self and continuity.
- Designing technologies that *support independence and autonomy* at home. In addition to using sensor-based technologies to create a safe environment for aging in place [31, 62], future work should explore the role of continuity in using technologies to support independence and autonomy at home, which would allow older adults to pursue meaningful activities without having to give up their interests due to certain constraints.
- Maintaining *consistency in digital interfaces and affordances.* To ensure continuity in older adults' competence with technology, future technology designs for older adults should use consistent languages and affordances and avoid mandating major updates where the interface changes significantly.

**Integrating Social Elements into IoT Systems.** From our participants' designs, we see an opportunity for future research on assistive technologies and social technologies to converge. Specifically, prior research on IoT systems for older adults has mainly focused on health monitoring [28], context sensing [63], reminder systems [61], and self-management [30]. While there are a few studies using ethnographic and co-design approaches to understand older adults' routines and home environments [1, 17], research on the integration of social elements into these systems is still limited. In our study, participants designed systems that integrated functions such as controlling household appliances with social functions such as calling family members and participating in online social activities. Our participants expressed a preference for integrated systems that provide streamlined experiences and flexible options for control. Future work could explore how integrated IoT systems can be designed to support older adults' meaningful social interaction at home.

**Using AI as Digital Coaches for Learning Support.** Our research suggests that participants need scaffolded support when using technology for meaningful activities. Some participants experienced difficulties when using the manuals of digital products or the help pages of some websites to solve problems and troubleshoot issues. Our participants envisaged the integration of AI into these systems as a way to provide interactive help and guidance. Recent researchers have been exploring the use of generative AI tools such as ChatGPT in the education field [44, 72]. These AI tools have the potential to be used as virtual intelligent tutors to provide personalised responses and feedback to students' questions [72]. Building on this, we see a broader opportunity to utilise AI as 'digital coaches' to support older adults' learning by providing step-by-step instructions and answering follow-up questions. For example, future work can explore how AI assistants can be used to support older adults' learning of feature-rich applications [59].

**Embracing the Complexity of Values in Technology Design.** Value-sensitive design (VSD) is an established method and theory in HCI research [15, 36, 66]. Building on this work, our study further highlights the important role of values in designing technologies that support older adults' engagement in meaningful activities. Notably, we found that while participants valued a range of activities that they found meaningful, they also valued safety and independence. We did not set out to investigate technology for safe ageing at home, yet our participants saw value in designing technologies that prioritised safety so as to enable continuous engagement in meaningful activities. Drawing on this finding, we argue that future researchers and designers should embrace the complexity of values in technology design. This entails considering not only human values but also values reflected in technology and the *nuanced interactions* between individuals and technologies. Waycott et al. [89] argued that the values embodied in technologies are sometimes inconsistent with the values of the people who come to use them, even after adopting a user-centred design approach. Previous research has identified a list of human and technology values that are important to older adults, such as independence [14, 40], privacy and security [19, 48], reciprocity [53], control and agency [13, 14], and competence [52]. Our study further reveals the spectrum of values and attributes that some older adults seek when using technologies to engage in meaningful activities, including simplicity, positivity, proactivity, and integration. Some of these values, particularly positivity and proactivity, are embedded in users' interactions with technology, influencing how they control the system and what feedback they receive from it. We believe that incorporating these nuanced values into technology design can foster deeper connections between older adults and the technologies they engage with.

### 5.3 Limitations and Future Work

A key limitation of this study is that we had a limited number of participants, all of whom were relatively healthy and experienced with technology and based in a metropolitan region in Australia. This means that some of the needs and values reflected in our study may be specific to the participants and may not be applicable to individuals with other life experiences. We also noted a gender imbalance in our participant group, with 12 women and only four men. We acknowledge that there could be more technologies, values, and tensions that are not covered in this study. Future research could be conducted with more diverse participant groups, including people with various cognitive functions and technological experiences. Nevertheless, we believe that our interpretation of the data, and particularly the application of Continuity Theory to our analysis, provides important insights for future researchers and designers. Our primary objective is not to generalise to the broader population. Rather, we aim to use the insights we learned to inform future technology design for others who wish to retain a sense of continuity as they transition into old age.

Additionally, our study centred on envisioned technologies and technology-based concepts rather than co-designing an established type of technology. We chose this approach to explore participants' desires and aspirations with maximum creativity and minimum realistic constraints. However, we also recognise the importance of future research addressing the practicality and feasibility of envisioned technologies. Future research should aim to strike a balance between conceptualisation and practical implementation, enhancing knowledge from various dimensions.

## 6 CONCLUSION

Engaging in meaningful activities is essential to the wellbeing of older adults and their quality of life. Through participatory workshops and interviews, our research reveals how participants interpreted the notion of meaningful activity and perceived the potential benefits and challenges of using emerging technologies to support such activities. We present how our participants envisioned future technologies for pursuing meaningful activities – they were concerned not only with creating technologies that can support or expand their current activities but also with maintaining their abilities and opportunities for continuing engagement in the future. These insights indicate a research direction focusing on the patterns of activity individuals develop in later stages of life and the role of digital technologies in supporting and maintaining these patterns, thereby contributing to a vibrant and active later life.

# APPENDICES

## A.1 Participant Demographics

| Participant ID | Age | Gender | Marital Status | Living Status | Self-described Technology Skill | Workshops Attended | Post-workshop interview |
|---|---|---|---|---|---|---|---|
| P1 | 71 | Man | Married | Living with partner | Advanced beginner | 1a | Yes |
| P2 | 71 | Man | Married | Living with partner | Advanced beginner | 1a, 2a | No |
| P3 | 69 | Woman | Married | Living with partner | Novice | 1a, 2a, 3 | Yes |
| P4 | 84 | Woman | Never married | Living alone | Advanced beginner | 1a, 2a, 3 | No |
| P5 | 81 | Woman | Never married | Living alone | Competent | 1a, 2a, 3 | Yes |
| P6 | 74 | Woman | Divorced | Living alone | Competent | 1a, 2a, 3 | Yes |
| P7 | 73 | Man | Widowed | Living alone | Competent | 1b, 2b, 3 | No |
| P8 | 67 | Woman | Never married | Living alone | Advanced beginner | 1a, 2a, 3 | Yes |
| P9 | 73 | Woman | Divorced | Living alone | Competent | 1b, 2b, 3 | Yes |
| P10 | 71 | Woman | Married | Living with partner | Competent | 1b, 2a, 3 | No |
| P11 | 78 | Woman | Widowed | Living alone | Advanced beginner | 1b, 2b, 3 | Yes |
| P12 | 74 | Man | Married | Living with partner | Advanced beginner | 1b, 2b | No |
| P13 | 72 | Woman | Married | Living with partner | Novice | 1b, 2b | No |
| P14 | 67 | Woman | Divorced | Living alone | Advanced beginner | 1b, 2b, 3 | Yes |
| P15 | 69 | Woman | Widowed | Living alone | Competent | 1b, 3 | Yes |
| P16 | 71 | Man | Married | Living with partner | Competent | 3 | Yes |

## A.2 Workshop 3 Design Topics

| No. | Design Topic | Short Title | Design Group |
|---|---|---|---|
| 1 | How can technologies be used to support your engagement in interests and hobbies (e.g., gardening, reading, cooking, arts)? | Interests and Hobbies | Group A |
| 2 | How can technologies be designed to support meaningful social interactions for you (e.g., connecting with family and friends, connecting with local communities)? | Meaningful Social Interactions | Group B |
| 3 | How can technologies help you stay physically and mentally active? | Staying Active | Group C |
| 4 | How can technology support you to live a meaningful life at home as you get older and avoid moving into care homes? | Aging at Home | Group D |
| 5 | How can we support older adults' building confidence in using technologies? | Building Confidence | |